\documentstyle[sprocl,rotate]{article}
\input{psfig.sty}
\def\be{\begin{equation}}
\def\ee{\end{equation}}
\def\bea{\begin{eqnarray}}
\def\eea{\end{eqnarray}}

\begin{document}
\title{NEUTRINO NUCLEUS SCATTERING}
\author{ PETR VOGEL }
\address{Physics Department 161-33, California
Institute of Technology, Pasadena, California 91125 \\
E-mail: vogel@lamppost.caltech.edu}
\maketitle
\abstracts{The status of the theoretical description of the neutrino-nuclear
interaction for low and intermediate energies is reviewed and its results are 
compared with the existing data. Particular emphasis is on $^{12}$C,
the ingredient of liquid scintillator, and on $^{16}$O,
the main component of the
water \v{C}erenkov detectors. First, 
I show that the data on the exclusive process
populating the ground state of $^{12}$N are well reproduced by the theory.
This is also the case for the excitation of the continuum with 
low energy neutrinos from the muon decay at rest and for muon capture. 
However, for not yet understood reasons, the theory overestimates the cross 
section for higher energy neutrinos from 
the pion decay in flight, by up to 50\%.
I also show that the Continuum Random Phase Approximation and the
Relativistic Fermi Gas model give very similar full and differential cross
sections for neutrino energies of 
several hundred MeV, thus checking one method
against the other. }
  
\section{Introduction and summary}

The observation of neutrinos and antineutrinos is often based on
their interaction with complex nuclei. 
The study of the corresponding
neutrino-nuclear cross section is rarely the primary motivation
of these experiments;
their aim most often belongs to the category of
``fundamental processes'', e.g., one tries to deduce from the outcome
either the properties of neutrinos themselves (as in the searches for
neutrino oscillations) or of the sources of the neutrinos (as in the attempts
to observe neutrinos from supernovae).

Neutrino-nuclear scattering is, therefore, for most people just a tool;
it belongs to ``neutrino engineering'' \footnote{I believe that
this, slightly derogatory but nevertheless rather accurate term was coined by
G. T. Garvey.}.
As with all tools, a solid understanding of cross sections 
in neutrino-induced
reactions, particularly on light nuclei, 
is a necessity. This includes especially
$^{12}$C, an ingredient of liquid scintillators, and $^{16}$O, the basic
component of water \v{C}erenkov detectors.  

There are numerous examples of application of nuclear targets. Let me
enumerate some of them, with brief comments:

\begin{itemize}
\item Detection of solar neutrinos.
This involves $\nu_e$ of the lowest energies in comparison with
essentially all other applications. When nuclear targets are used (chlorine, gallium,
iodine, etc.) usually bound discrete final states are populated.

\item Signal in the KARMEN and LSND detectors.
The sources involve somewhat higher energy neutrinos
(e.g. from the pion and muon decays at rest). The target
is liquid scintillator containing hydrogen and $^{12}$C nuclei. As I will
describe in more detail later, neutrinos excite both the discrete bound states
and the continuum final states.

\item Detection of atmospheric neutrinos. 
Here the energies are substantially higher. In most applications the
target nuclei are $^{16}$O (water \v{C}erenkov detectors). Essentially all
final states are in the continuum. 
Details of nuclear structure play a secondary,
but still nonnegligible role.

\item r-process nucleosynthesis.
There, neutrinos of all flavors and moderate energies are expected to
interact with nuclei far from stability.  Both discrete and continuum final
states are important. This application of 
the neutrino-nuclear interaction is based,
at present, only on theoretical estimates of the corresponding cross sections.

\item Detection of supernova $\nu_{\mu}$ and $\nu_{\tau}$. These neutrinos
will interact only through the neutral current, and can be detected by
e.g. observation of the deexcitation of $^{16}$O in water \v{C}erenkov
detectors. Again, the corresponding cross section is based only
on theoretical estimates.

\end{itemize}

Since only a few of the just enumerated cross sections have been measured,
one has to rely often on nuclear theory for their evaluation. i
The description of the
theoretical effort, and comparison with existing data, are the topics of this
talk.

Ideally, one would like to have a relatively 
simple universal recipe valid for all
nuclei and all neutrino energies. Alas, but perhaps not surprisingly,
my main message is that such a recipe
does not exist. Instead, for each energy one needs a 
somewhat different approach.
These approaches then should smoothly connect at the corresponding 
boundaries of applicability.

\begin{itemize}
\item At the lowest energies the details of nuclear structure, 
i.e., all the complications
related to the many-body nature of the nuclear system, 
really matter. The nuclear
{\em shell model} is then the method of choice.

\item At intermediate energies the shell model 
becomes untractable (or essentially so).
At the same time, the particle-hole nature of the final nuclear state becomes
the most important feature. Some form of 
the {\em Random Phase Approximation} (RPA)
 is then the method of choice. 
 
\item At yet higher energies, 
the details of the nucleon-nucleon interaction become
less important. The {\em Fermi gas model}  is then the method of choice.
 
\end{itemize}
 
 In the following I will concentrate on the 
neutrino-$^{12}$C interaction where a number
of experimental results exist. These include
measurements of charged-current reactions induced by both
electron- \cite{Karmen,LSND} and muon-neutrinos \cite{LSND},
exciting both the ground and continuum states in $^{12}$N.  The inclusive
cross section for $^{12}$C($\nu_e,e$)$^{12}$N$^{*}$ 
\cite{Karmen,LSND,Krakauer},
agrees well with calculations. 
By contrast, there is a discrepancy between
calculations\cite{Kolbe94,Kolbe95,Oset,Oset98} 
(with some notable exceptions \cite{Mintz,Auerbach}) 
and the measured\cite{LSND} inclusive cross
section for  $^{12}$C($\nu_\mu,\mu$)$^{12}$N$^{*}$,
which uses higher energy neutrinos from pion decay-in-flight.  
The disagreement is disturbing 
in light of the simplicity of the reaction
and in view of the fact that parameter-free calculations, such as
those in\cite{Kolbe94,Kolbe95}, describe well other weak processes
governed by the same weak current nuclear matrix elements. 
The exclusive reactions populating the ground state of the final nucleus,
$^{12}$C($\nu_e,e$)$^{12}$N$_{gs}$ and
$^{12}$C($\nu_\mu,\mu$)$^{12}$N$_{gs}$,
and the neutral current reaction $^{12}$C($\nu_e,\nu_e '$)
$^{12}$C(15.11 MeV) have been measured\cite{Karmen,LSND} as well,
and agree perfectly with theoretical expectations.

My own theoretical work, which I will review below,
is based on a series of calculations performed in collaboration
with Edwin Kolbe, Karlheinz Langanke and Jonathan Engel.

\section{Exclusive Reactions}
 
Among the states in the final nucleus $^{12}$N,
which is populated by the charged current reactions 
with beams of $\nu_e$ or $\nu_{\mu}$, 
the ground state $I^{\pi} = 1^+$
plays a special role. It is the only 
bound state in $^{12}$N, and can be recognized
by its positron decay ($T_{1/2} = $ 11 ms) back to $^{12}$C.
Moreover, the analog of the $^{12}$N$_{gs}$, the $I^{\pi} = 1^+$ state 
with isospin $T = 1$ at 15.11 MeV in $^{12}$C, can be populated by the 
neutral current neutrino scattering, and is recognizable by its emission
of the 15.11 MeV photon. Finally, even though there are several bound
states in $^{12}$B,  its ground state, the analog of the other two 
$I^{\pi} T= 1^+1$ states, is the state most strongly populated
in muon capture on $^{12}$C. Again, the population of the bound states 
in $^{12}$B can be separated from the continuum by observing its electron 
decay ($T_{1/2} = $ 20.2 ms). 

\begin{table}[h]
\caption {\protect Comparison of calculated and measured cross sections, in 
units of $10^{-42}$cm$^{-2}$ and averaged over
the corresponding neutrino spectra, for the neutrino induced transitions 
$^{12}$C$_{gs} \rightarrow ^{12}$N$_{gs}$ and $^{12}$C$_{gs} \rightarrow 
^{12}$C(15.11 MeV). For the decay at rest the $\nu_e$ spectrum is normalized
from $E_{\nu}$ = 0, while for the 
decay in flight the $\nu_{\mu}$ and $\bar{\nu}_{\mu}$
spectra are normalized from the corresponding threshold.
See the text for explanations.}
\label{tab:gs}
\vspace{0.4cm}
\begin{center}
\begin{tabular}{|lccc|} 
\hline
  & $^{12}$C($\nu_e,e^-)^{12}$N$_{gs}$  & 
$^{12}$C($\nu_{\mu},\mu^-)^{12}$N$_{gs}$  &
 $^{12}$C($\nu,\nu')^{12}$C(15.11)  \\
  &  decay at rest   & decay in flight  & decay at rest \\
\hline  
experiment\cite{Karmen} & 9.4$\pm0.5\pm0.8$ &  - &11$\pm$0.85$\pm1.0$ \\
experiment\cite{LSND} & 9.1$\pm0.4\pm0.9$  & 66$\pm10\pm10 $ & - \\
experiment\cite{Krakauer} & 10.5$\pm1.0\pm1.0$ &  - & - \\
Shell model \cite{triad} & 9.1 & 63.5  & 9.8 \\
CRPA \cite{Kolbe94,Kolbe95} & 8.9 & 63.0 & 10.5 \\
EPT \cite{Fukugita} & 9.2 & 59 &  9.9 \\
\hline
\end{tabular}
\end{center}
\end{table}

\begin{figure}
\centerline{
\psfig{figure=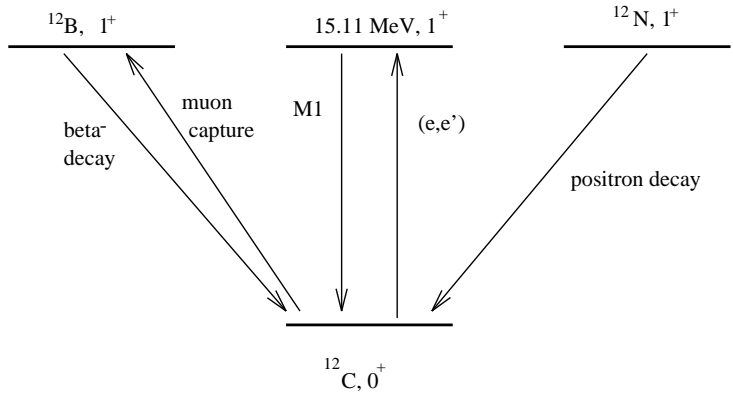,height=2.0in}}
\caption{Constraints on the exclusive channels.
\label{fig:excl}}
\end{figure}

Theoretical evaluation of the exclusive 
cross sections is constrained by the obvious
requirement that the same method, and the same parameters, must also describe
the related processes, shown schematically 
in Fig. \ref{fig:excl}. It turns out that this
requirement essentially determines the neutrino induced cross section for
the energies of present interest. It does not 
matter which method of calculation
is used, as long as the constraints are obeyed.

The comparison between the measured and calculated values is shown in 
Table \ref{tab:gs}. There, three rather different methods of calculation
were used, all giving excellent agreement with the data.

The first approach is a restricted shell-model calculation.
Assuming that all structure in the considered low-lying states 
is generated by the  
``valence" nucleons in the 
$p$-shell, there are only four one-body densities (OBD)
which fully describe all necessary nuclear matrix elements.
The most straightforward way of obtaining the OBD is by
diagonalizing a thoroughly tested residual interaction.  
However, the resulting $p$-shell OBD
do not describe the processes in Fig. \ref{fig:excl} very well; 
to remedy this one can
modify the one-body densities (ad hoc) in such a way that all these
``auxiliary'' data are correctly reproduced. This then gives the
results listed in line 4 of Table \ref{tab:gs}.

Effects of configurations beyond the $p$ shell might explain the need
for the renormalization of the one-body densities produced by a
reasonable $p$-shell Hamiltonian.  We therefore also calculate the rates of all
the reactions above, including exclusive neutrino capture, in the Random
Phase Approximation (RPA), which does include multishell correlations, while
treating the configuration mixing within the $p$ shell only crudely.
Again an adjustment is needed (a ``quenching'' of all matrix elements by
an universal, but substantial, factor 0.515). 
However, the neutrino cross sections
in line 5 of  Table \ref{tab:gs} agree with the measurements perfectly.

The third approach is the ``elementary-particle treatment" (EPT).  Instead of
describing nuclei in terms of nucleons, the EPT
considers them elementary and describes transition matrix elements in
terms of {\it nuclear} form factors deduced from experimental data.  
The EPT approach was extended in Ref. \cite{triad} to the
higher neutrino energies relevant to the LSND decay-in-flight $\nu_{\mu}$'s
by appropriately including the lepton mass.

\begin{figure}
\centerline{
\psfig{figure=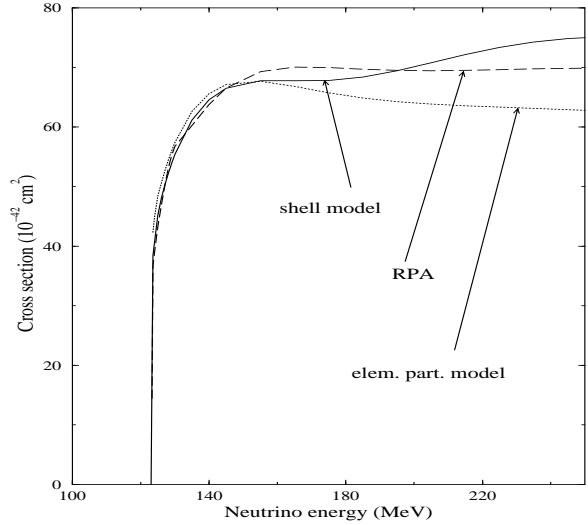,height=3.0in,width=3.0in}}
\caption{Energy dependence of the cross section for the
reaction 
$^{12}$C + $\nu_{\mu} \rightarrow ^{12}$N$_{g.s} + \mu^-$.
\label{fig:gs}}
\end{figure}

An example of the energy dependence of the exclusive cross section
is shown in Fig. \ref{fig:gs} for the $\nu_{\mu}$ induced exclusive reaction.
As one can see, the cross section raises sharply from its threshold
($E_{thr} = $ 123 MeV) and soon reaches its saturation value, i.e.,
it becomes almost energy independent. This means that the yield of
the$^{12}$C + $\nu_{\mu}$ reaction essentially measures just the
flux normalization above the reaction threshold. At the same time,
the yield is insensitive to the energy distribution of the muon
neutrinos in the beam.

\section{Inclusive Reactions}

The inclusive
reactions $^{12}$C($\nu_e,e$)$^{12}$N$^{*}$, 
with $\nu_e$ neutrinos from the muon decay-at-rest
and $^{12}$C($\nu_\mu,\mu$)$^{12}$N$^{*}$
with the higher energy $\nu_{\mu}$ neutrinos from the pion
decay-in-flight populate not only the ground
state of $^{12}$N but also the continuum states.
The corresponding cross sections involve folding over
the incoming neutrino spectra and integrating over the
excitation energies in the final nucleus.
By convention, we shall use the term ``inclusive'' for the
cross section populating only the continuum (i.e., without 
the exclusive channel) for $^{12}$C($\nu_e,e$)$^{12}$N$^{*}$
with the decay-at-rest $\nu_e$, while for the reaction
$^{12}$C($\nu_\mu,\mu$)$^{12}$N$^{*}$ with the decay-in-flight
$\nu_{\mu}$ the term is used for the total cross section
(the exclusive channel then represents only a small fraction
of the total).

\begin{figure}
\rotate[l]{
\psfig{figure=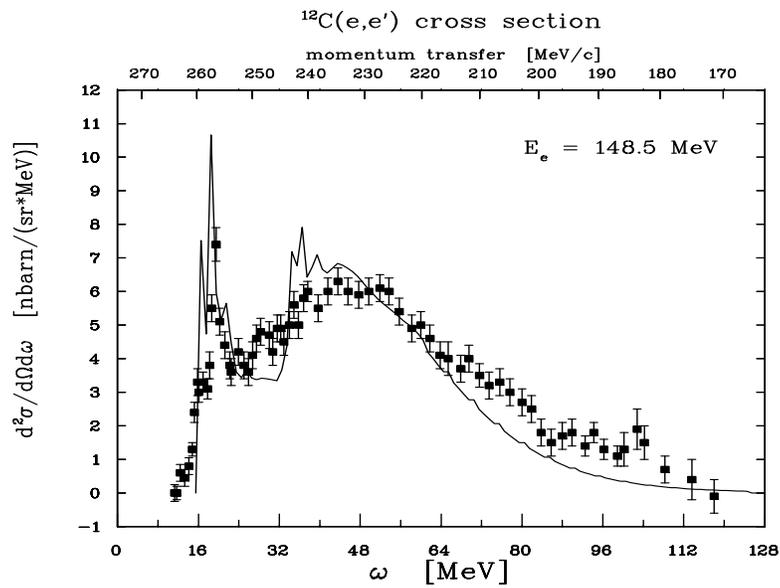,height=4.0in,width=3.0in}}
\caption{ Data (points with error bars) and calculated cross section
for the inclusive electron scattering on $^{12}$C as a function of the
excitation energy $\omega$. The corresponding momentum transfer
is displayed on the upper scale.
\label{fig:elsc}}
\end{figure}

Muon capture, $^{12}$C($\mu, \nu_{\mu}$)$^{12}$B$^{*}$,
belongs also to this category. It involves momentum transfer
of $q \approx m_{\mu}$, intermediate between the two neutrino
capture reactions above. Since $^{12}$B and $^{12}$N are
mirror nuclei, all three reactions should be considered
together. In this case again the term ``inclusive'' will be used
only for the part of the rate populating the continuum in $^{12}$B.

What theoretical approach should one use in order to describe
such reactions?  One possibility is to use the  
continuum random phase approximation (CRPA). The method has been
used successfully in the evaluation of the nuclear response to
weak and electromagnetic probes \cite{Speth}. In particular, we have
tested it, with good agreement, in the calculation of the inelastic electron
scattering\cite{Leiss} on  $^{12}$C involving very similar excitation energies 
and momentum transfers as the weak processes of interest.
As an example I show in Fig. \ref{fig:elsc}  the comparison of the
experimental data and the results of the CRPA for the inclusive
electron scattering\cite{antimintz}. One can see that CRPA describes quite 
well both the magnitude and shape of this cross section over the
entire range of excitation energies and momentum transfers.

For muon capture the CRPA \cite{capture} gives the inclusive rates of 
0.342, 0.969, and 26.2 $\times 10^5$ s$^{-1}$ for $^{12}$C, $^{16}$O
and $^{40}$Ca; to be compared with the measured rates of
0.320, 0.924, and 25.6 $\times 10^5$ s$^{-1}$ for the same nuclei.
This good agreement is again obtained without any parameter adjustment.
In particular, as discussed in Ref.\cite{capture}, no renormalization of the
axial vector coupling constant $g_A$ in nuclear medium is required.

Can one understand why CRPA is apparently able to describe the inclusive
processes without the need for parameter adjustment, unlike the case of the
exclusive reactions discussed earlier? Another way of thinking about this
problem is the question to what extent the correlations of nucleons within the
nuclear $p$ shell influence the result. 
In order to shed light on this, one can evaluate the total strength of various
operators $\hat{O}$, i.e. the norm of the vector 
$\hat{O}|g.s.\rangle$, with and without
the effect of the correlations. 
Note that the total strength depends only on the ground state
wave function and is therefore relatively easy to evaluate.
For the positive parity operators this is done
in Table \ref{tab:fs}. 

\begin{table}[h]
\caption {\protect  The full strength within the nuclear
$p$ shell evaluated for the operators in column 1.
The SM column is the exact shell model strength calculated with the
Cohen-Kurath interaction. The ``naive'' column corresponds to the
$(1p_{3/2})^8$ configuration, i.e. no correlations whatsoever. In  the
last column the strength for transitions with $2 \hbar \omega$ excitation
energy is shown for comparison.}
\label{tab:fs}
\vspace{0.4cm}
\begin{center}
\begin{tabular}{|lccc|} 
\hline
Operator & SM & ``naive'' & $2 \hbar \omega$  \\
\hline  
GT$(\sigma \tau)$ & 1.51 & 8.00 & 0.00 \\
$r^2 Y_2$ & 1.37 & 1.98 & 9.95 \\
$r^2 (Y_2 s)^{I=1}$ & 0.11 & 0.08 & - \\
$r^2 (Y_2 s)^{I=2}$ & 0.33 & 0.75 & - \\
$r^2 (Y_2 s)^{I=3}$ & 0.20 & 0.00 & - \\
$ \sum_{\lambda} r^2 (Y_2 s)^{I=\lambda}$ & 0.63 & 0.83 & 7.47 \\
\hline
\end{tabular}
\end{center}
\end{table}

Table \ref{tab:fs} illustrates the well
known fact that the $p$ shell correlations are very
important for the Gamow-Teller operator $\sigma \tau$.
(Note that RPA gives the total GT strength of 5.5, only slightly reduced
when compared to the ``naive'' estimate. 
Also, the exact shell model predicts that
a strength of 0.12 goes to excited $1^+$ states, which are absent when the
naive model is considered.)

\begin{figure}
\centerline{
\psfig{figure=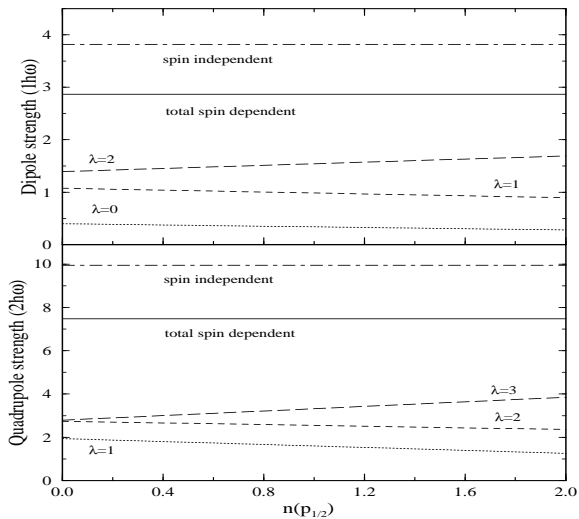,height=2.7in,width=3.0in}}
\caption{ Total strength of the dipole (upper part) and quadrupole (lower
part) operators versus the occupation of the $p_{1/2}$ subshell. The
curves are labelled by the corresponding angular momentum of
the $r^l (Y_l s)^{I=\lambda}$ operators for $l =1$ and $l=2$.
\label{fig:str}}
\end{figure}

But the situation with quadrupole operators is rather different. The total
$p$ shell strength of the spin-independent operator, and the strength
summed over the multipolarities of the spin-dependent operators
is affected by the correlations only
at the level of 30-40\%, even though the individual spin dependent
multipoles are affected more. Moreover, the $p$ shell strength represents
only a small fraction of the total 
quadrupole strength, which is concentrated in the
$2 \hbar \omega$ excitations, unaffected by correlations as long as
we assume that the ground state has only $p$ shell nucleons.

However, the inclusive reactions we are considering are dominated by the
excitations of the negative parity states. 
To what extent does the strength depend on
the occupation of the $p_{1/2}$ subshell? 
Note that Auerbach et al.\cite{Auerbach} 
claim that when there are about 1.6 nucleons in 
the $p_{1/2}$ subshell, the inclusive cross section is reduced
substantially. In order to test the sensitivity to this
``pairing'' effect, we plot in Fig. \ref{fig:str} the dipole and quadrupole
strengths as a function of the $p_{1/2}$ subshell occupation. We find
that when summed over multipoles the strength is totally independent
of this occupation number. But even the individual multipoles depend
on the occupation numbers only mildly. Thus, at least for the full strength,
we find again that $p$ shell correlations are relatively unimportant.

What are the momentum transfers and excitation energies
involved in the inclusive reactions which we would like
to describe? For the $^{12}$C($\nu_e,e$)$^{12}$N* with the
electron neutrinos originating in the muon decay at rest,
the typical momentum transfer is $\langle |\vec{q}| \rangle \simeq$ 
50 MeV, and the typical excitation energy
is $\omega \simeq$ 20 MeV. For the inclusive muon
capture $^{12}$C($\mu^-,\nu_{\mu}$)$^{12}$B* we have
$\langle |\vec{q}| \rangle\simeq$ 90 MeV 
and the typical excitation energy
is $\omega \simeq$ 25 MeV. Finally for the 
$^{12}$C($\nu_{\mu},\mu^-$)$^{12}$N* 
with the muon neutrinos originating in the pion decay
in flight at LAMPF we have 
$\langle |\vec{q}| \rangle\simeq$  200 MeV 
and the typical excitation energy
is $\omega \simeq$ 40 MeV.
The excitation energies should be compared with the
nuclear shell spacing $\hbar \omega \simeq 41/A^{1/3}$ MeV, which for
$^{12}$C is equal to about 18 MeV. Thus, in order to describe 
all the above inclusive processes in the framework of the nuclear
shell model, one would have to include fully and consistently at
least all $2 \hbar \omega$ excitations, and possibly even
the $3 \hbar \omega$ ones. That is not impossible, but 
represents a formidable 
task. On the other hand, CRPA can easily handle such configuration
spaces. Moreover, it properly describes the continuum nature
of the final nucleus. Finally, as argued above, the crudeness with which
the correlations of the $p$ shell nucleons is treated in the CRPA
is expected to be relatively unimportant.
 
For the inclusive reaction $^{12}$C($\nu_e,e^-$)$^{12}$N$^{*}$, 
with $\nu_e$ neutrinos from the muon decay-at-rest
the calculation gives\cite{Kolbe94} 
the cross section of 6.3 $\times 10^{-42}$ cm$^2$
using the Bonn potential based G-matrix as the residual interaction,
and 5.9 $\times 10^{-42}$ cm$^2$ with the schematic Migdal force.
(The two different residual interactions are used so that
one can estimate the uncertainty associated with this aspect
of the problem.) 
Both are clearly compatible with the measured values of
$6.4 \pm 1.45[stat] \pm 1.4[syst] \times 10^{-42}$ cm$^2$  by
the Karmen collaboration\cite{Karmen} (the more
recent result gives somewhat smaller value
$5.1 \pm 0.6 \pm 0.5$\cite{Maschuw}) and with
$5.7 \pm 0.6[stat] \pm 0.6[syst] \times 10^{-42}$ cm$^2$ obtained by
the LSND collaboration\cite{LSND} . If one wants to disregard
the error bars (naturally, one should not do that),
one can average the two calculated values
as well as the two most recent measurements and
perhaps conclude that the CRPA calculation
seems to exceed the measured values by about 10-15\%. 
A similar tendency can be found, again with some
degree of imagination, in the comparison of the muon
capture rates discussed earlier.

So far we have found that CRPA describes the
inclusive reactions quite well. Other theoretical calculations,
e.g.\cite{Oset98,Auerbach} describe these reactions
with equal success. This is no longer the case when we
consider the reaction $^{12}$C($\nu_\mu,\mu$)$^{12}$N$^{*}$
with the higher energy $\nu_{\mu}$ neutrinos from the pion
decay-in-flight. This reaction involves 
larger momentum transfers and populates
states higher up in the continuum of  $^{12}$N. 
Our calculation\cite{Kolbe94,Kolbe95} gives the cross section of 
19.2 $\times 10^{-40}$ cm$^2$,
considerably larger than the measured\cite{LSND} value of 
$11.3 \pm 0.3[stat] \pm 1.8[syst]$ in the same units.
The origin of the discrepancy is not clear, but as stressed in the discussion
of the exclusive reaction, the $\nu_{\mu}$ flux normalization is not
a likely culprit. While Ref. \cite{Oset} confirms our result, Ref.\cite{Mintz}
gets a value close to the experiment by using a generalization of the
EPT approach. It is questionable, however, that the assumptions 
used in\cite{Mintz}
are justified\cite{antimintz}.

Other recent theoretical calculations span the region between
the CRPA and experiment. So, Singh et al.\cite{Oset98} give
$16.65 \pm 1.37$ $\times 10^{-40}$ cm$^2$, clearly higher
than the experiment but somewhat lower than the CRPA.
On the other hand, Ref.\cite{Auerbach} gives 13.5 - 15.2 in the
same units, value which is even closer to the experiment. 
The main difference in that work is the inclusion of pairing, which
as I argued above, should not represent a substantial effect.

This discrepancy has been with us for quite some time now.
A small, but not insignificant step which removes 
part of it
is the new simulation of the $\nu_{\mu}$ flux for the
decay-in-flight beam (R. Imlay, private
communication). As I emphasized above, the exclusive reaction
fixes to some extent the flux normalization, but is insensitive
to its shape. The revision, which is within the error bars of the previous
flux, results in moving the most probable measured value up
from 11.2 to 12.4 (subject to revision and with as yet undetermined
error bars). At the same time, the CRPA calculated value moves
down from 19.3 to 18.0 (all in  $10^{-40}$ cm$^2$).
While diminished, the discrepancy is still clearly present, 
and also clearly exceeds the 10-15\% perhaps suggested by the
lower energy inclusive reactions discussed above.
It would be very important to perform a large scale shell model
calculation, including up to 3$\hbar \omega$
excitations,  to put the matter to rest.

The importance of the inclusive $^{12}$C($\nu_{\mu},\mu^-$)$^{12}$N$^{*}$
reaction goes beyond its significance for testing our ability
to perform calculations of this kind. The LSND collaboration
announced evidence for $\nu_{\mu} \rightarrow \nu_e$ oscillations
based on the same $\nu_{\mu}$ neutrino beam from the pion decay
in flight \cite{LSNDp}. The resulting electron neutrinos $\nu_e$
are detected by their charged current interaction with the $^{12}$C
nuclei, and are recognized by the observation of high energy 
(60 - 200 MeV) electrons. In order to extract the oscillation
probability from the observed number of events, however,
one has to know 
the corresponding inclusive cross section, analog of which
is the discrepant result just discussed.

\begin{figure}
\centerline{
\psfig{figure=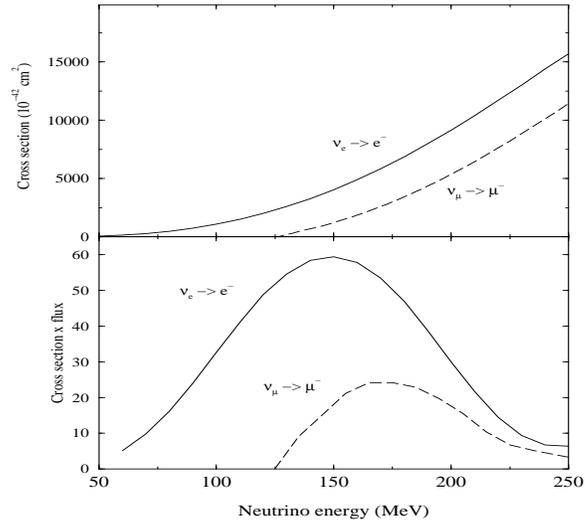,height=3.0in,width=3.0in}}
\caption{The cross sections for indicated reactions on $^{12}$C 
are compared in the lower panel, while in the upper panel the
cross section is multiplied by the corresponding decay-in-flight flux.
\label{fig:comp}}
\end{figure}

One has to remember, however, that  
the $\nu_{\mu}$ induced reaction has a threshold of 123 MeV,  
while the $\nu_e$ induced reaction has a threshold of only 17 MeV
(or with the experimental contraint on the electron energy the 
effective threshold
is about 80 MeV). At the same time the decay-in-flight
neutrino beam is essentially monotonically decreasing
with the neutrino energy. Consequently the neutrino
energies involved in the $\nu_e$ induced reaction
will be smaller than in the $\nu_{\mu}$ induced reaction.
This is illustrated in Fig. \ref{fig:comp}
where cross sections for both reactions are compared in 
the upper panel. In the lower panel, assuming
full conversion  $\nu_{\mu} \rightarrow \nu_e$, I plot
the cross section$\times$neutrino flux (i.e. the number of events)
as a function of the neutrino energy. 
One can clearly see that the oscillation signal would be caused by
neutrinos of lower energy on average than the neutrinos
involved in the inclusive $^{12}$C($\nu_{\mu},\mu^-$)$^{12}$N$^{*}$
reaction. While this does not guarantee that the uncertainty in the
cross section will be lower, it makes this assumption rather plausible.

\section{Angular distribution: CRPA versus Fermi gas model}

\begin{figure}
\centerline{
\psfig{figure=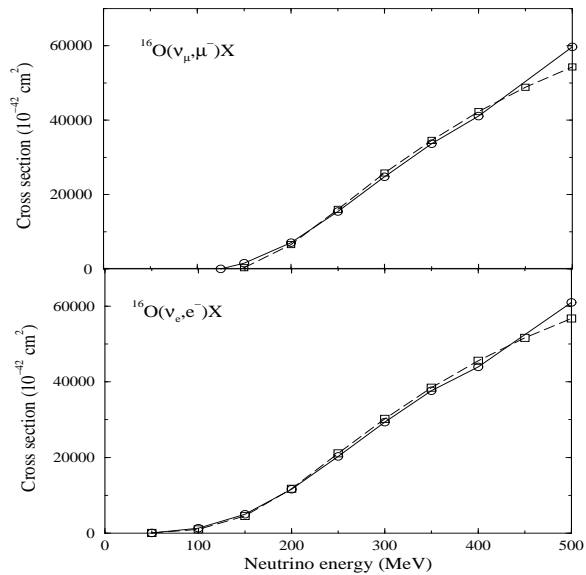,height=3.0in,width=3.0in}}
\caption{The cross sections for indicated reactions on $^{16}$O
in units of $10^{-42}$ cm$^{-2}$. The dashed line is for the relativistic
Fermi gas method and the full  line is for the continuum
random phase approximation (CRPA).
\label{fig:o16t}}
\end{figure}

We now turn our attention to the neutrino induced reactions at higher energies,
closer to those encountered in the study of atmospheric neutrinos. 
As we stressed earlier, at sufficiently high neutrino energy the description
using the CRPA method and the relativistic Fermi gas (RFG) should give
identical, or at least
very similar results. Here we would like to compare the two
methods not only as far as the full cross section is concerned, but also for the
description of the angular distribution of the emitted lepton.
We use the Smith and Moniz formulation of the RFG method
\cite{Moniz}. Earlier, the CRPA has been applied to the atmospheric
neutrino problem in Ref. \cite{atm}. 

\begin{figure}
\centerline{
\psfig{figure=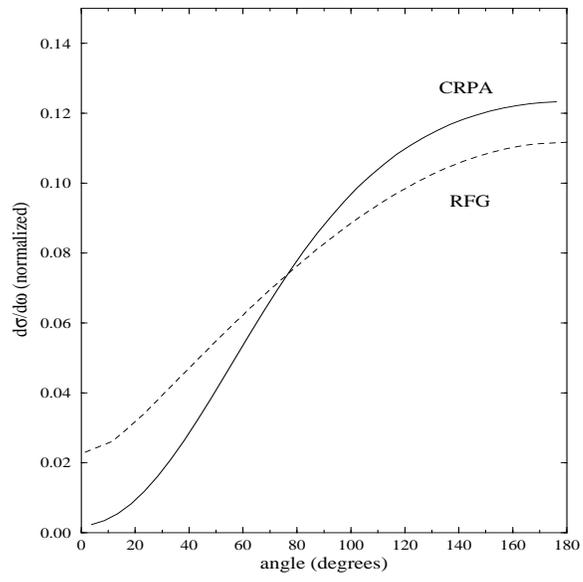,height=3.0in,width=3.0in}}
\caption{The differential cross sections for 
the reaction $^{16}$O$(\nu_e, e^-)X$ for the neutrino
energy $E_{\nu}$ = 100 MeV.
The dashed line is for the relativistic
Fermi gas method and the full line is for the continuum
random phase approximation (CRPA).
\label{fig:o100}}
\end{figure}

The application of CRPA to the angular distribution in the neutrino 
induced reactions was developed by Kolbe \cite{Kolbe96} who applied
it to the charged current reactions on $^{12}$C relevant to the LSND
experiment. A satisfactory agreement with the experimental  distributions
was obtained. The results presented here were also obtained by Kolbe 
and will be reported on in detail elsewhere.

In Fig. \ref{fig:o16t} I show the total cross section for the charged current
reactions on $^{16}$O calculated by both methods for neutrino energies
up to 500 MeV. One can see that over that energy interval the two methods
agree quite well. (The RFG was evaluated with the standard values $p_f$
= 225 MeV  and $e_b$ = 27 MeV deduced from the electron-nucleus
scattering data \cite{elsc}.)

The problem of the angular distribution is very important these days.
The Super-Kamiokande collaboration reported evidence for oscillation
of atmospheric neutrinos \cite{SK}. The data show a zenith angle dependence
of the muon neutrino deficit for both the sub-GeV and multi-GeV samples.
Of particular interest for the present work is the observation
that the muons are essentially isotropic for momenta $p < $ 400 MeV,
and show a zenith angle anisotropy at higher momenta.

\begin{figure}
\centerline{
\psfig{figure=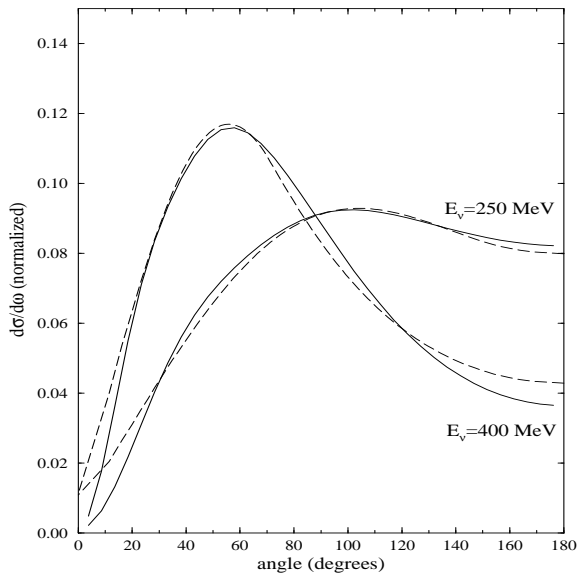,height=3.0in,width=3.0in}}
\caption{The differential cross sections for 
the reaction $^{16}$O$(\nu_e, e^-)X$ for the neutrino
energies of $E_{\nu}$ = 250 and 400 MeV.
The dashed line is for the relativistic
Fermi gas method and the full line is for the continuum
random phase approximation (CRPA).
\label{fig:o250}}
\end{figure}

Of course, the observed direction 
in Super-Kamiokande is the direction of the 
muon or electron. The direction
of the neutrino is then deduced from the expected angular correlation in the
neutrino - oxygen quasi-elastic charged current reaction.
The analysis in Ref.\cite{SK} is based on RFG. How good is that for
neutrino energies of a few hundred MeV?

First, as pointed out by Kolbe\cite{Kolbe96} at energies
corresponding to the neutrino beams available  at LAMPF
in both the $^{12}$C($\nu_{\mu},\mu^-$)$^{12}$N$^{*}$ 
and $^{12}$C($\nu_e, e^-$)$^{12}$N$^{*}$ reactions
the muons and electrons are in fact backward peaked.
This is also true for the charged current reactions on
$^{16}$O as shown in Fig. \ref{fig:o100}. There one can see that
both methods predict backward peaking which is considerably
more pronounced in CRPA than in the RFG method.

In Fig.\ref{fig:o250} I compare the differential
cross section for the two methods for neutrino energies
of 250 and 400 MeV. I use here electrons as outgoing leptons
so that the outgoing particle is fully relativistic, and the 
threshold effects are negligible. One can see in Fig. \ref{fig:o250}
that, first of all, at these energies the two methods give 
essentially identical results, as expected. Also, the peak in angular
distribution is gradually shifting to forward angles. That is
again in at least qualitative agreement with the
Super-Kamiokande finding where very 
little asymmetry was found for muons with momenta less
than 400 MeV. 

The results of this Section thus serve two purposes. First, they show
that at energies of several hundred MeV the continuum random phase
approximation (CRPA), 
and the relativistic Fermi gas (RFG) model give essentially
identical results both for the full and differential cross sections.
Let me stress that both calculations are parameter free, in the sense
that the parameters involved were obtained independently, not adjusted
for the weak processes in question.
The second lesson, related to the first one, is an indirect confirmation
of the procedure used to analyze the zenith angle dependence of the
atmospheric neutrino anomaly.   

\section{Yet higher energies}

In this section I would like to report on a piece of
``work in progress''. So far, I have shown that at
the lowest energies, corresponding to the
muon decay-at-rest neutrinos, various methods
including the CRPA and the nuclear shell model,
give very similar results. This then tests the one method 
against the other. Also, at those low neutrino energies,
the {\em absolute} cross sections are correctly
reproduced.

\begin{figure}[ht]
\centerline{
\psfig{figure=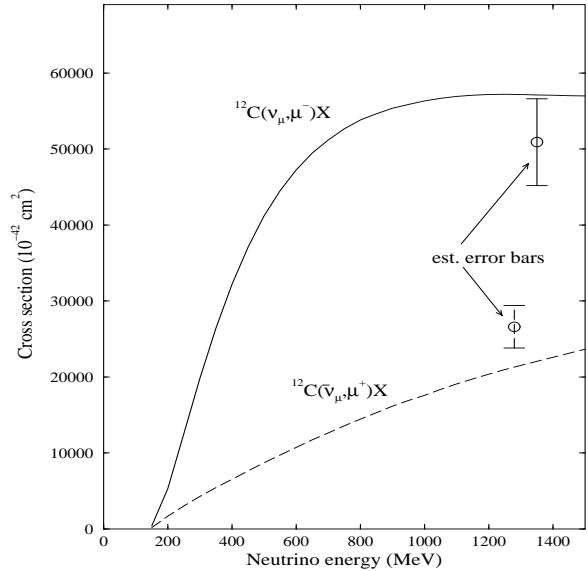,height=3.0in,width=3.0in}}
\caption{Cross sections for the indicated reactions
calculated using the RFG model with parameters determined
from the quasielastic electron scattering ($e_F =$ 221 MeV,
$e_b =$ 25 MeV). The Brookhaven AGS data with estimated 
error bars, and for an average energy, are shown for comparison.
\label{fig:bnl}}
\end{figure}

Next came the discussion of the inclusive reactions using
the pion decay-in-flight $\nu_{\mu}$ of somewhat higher
energy. There, unfortunately, reliable shell model calculations
are not available yet. At the same time, the CRPA overestimates
the measured cross section by about a factor 1.5. Even though
various theoretical papers come a bit closer to the experiment,
those calculations have not been so throughly tested as the CRPA.
In any case, a majority of these results also gives cross sections
which are larger than the measured one. The reasons for this 
discrepancy remain unclear. However, it is possible that the problem
lies in our ability to describe the neutrino-nucleus interaction at
these energies.

Next, in the previous section, I have shown that at energies of
several hundred MeV the CRPA and RFG give essentially identical
total and differential cross sections. The obvious experimental
data at these energies come from the study of atmospheric neutrinos,
and are subject to the uncertainties associated with the incoming
flux of the atmospheric neutrinos.
Thus it is difficult to make a model independent comparison between
the calculated and measured absolute cross sections.

However, at yet higher energies, of about 1 GeV, the cross section
for $\nu_{\mu}$ and $\bar{\nu}_{\mu}$ has been measured in a series
of experiment at the Brookhaven AGS in the early eighties 
(see e.g. \cite{Ahrens87,Ahrens88}). These experiments, among other
things, allowed the determination of 
the axial vector form factor of the nucleon.
The detector used at AGS was a liquid scintillator based detector\cite{NIM};
the cross section, measured for both $\nu_{\mu}$ and $\bar{\nu}_{\mu}$ beams
thus represent the quantity we have been discussing all along, namely
the cross section for the neutrino-$^{12}$C quasielastic scattering.
Thus it is tempting to extend the RFG model calculation to the
corresponding energies, and compare the calculated and measured
cross sections. One would be able to see in that way whether the agreement
is restored again at these energies where nuclear structure presumably
plays relatively minor role.

The problem is the extraction of the absolute cross section from the published 
data. The quasielastic scattering cross section was used in the experiment
to verify the Monte-Carlo simulation of the beam. From the consistency of
the measured and expected rates it is obvious that the modeling of the beam
was correct, but it would be desirable to make this statement more quantitative.
This is the part of the analysis that remains to be done.

The preliminary comparison is presented in Fig. \ref{fig:bnl} for both beams. 
The experimental error bars are just estimates, as explained above. Also, the
wide band beam is replaced by a beam with single average energy. (That is not
too critical at this stage, since the  cross section almost saturates at the 
considered energies.) With all these caveats, Fig. \ref{fig:bnl} shows that the
RFG describes the data to accuracy not worse than about 20\%, and likely
much better. Whatever is causing the discrepancy at the LAMPF energies
has healed itself, as expected, at these much higher energies.

\section*{Acknowledgments}

It is my pleasant duty to thank my collaborators, 
Jon Engel, Karlheinz Langanke,
and especially Edwin Kolbe. I have also benefited 
from discussions with members
of the LSND collaboration, in particular with 
R. Imlay, W. Louis, and D. H. White who
inspired me to consider the higher energy extension of the present calculation.
This work was supported by the US Department of Energy under Grant No.
DE-FG03-88ER-40397.

\end{document}